**Astro2020 Science White Paper**

**Reconstructing Extreme Space Weather from Planet Hosting Stars**

**Thematic Areas:** ☒ Planetary Systems ☐ Star and Planet Formation
☐ Formation and Evolution of Compact Objects ☐ Cosmology and Fundamental Physics
☒ Stars and Stellar Evolution ☐ Resolved Stellar Populations and their Environments
☐ Galaxy Evolution ☐ Multi-Messenger Astronomy and Astrophysics

**Principal Author**:
Name: Vladimir Airapetian
Institution: NASA GSFC/SEEC and American University, DC
E-mail: vladimir.airapetian@nasa.gov

**Co-authors**: V. Adibekyan (UPORTO), M. Ansdell (University of Hawaii), D. Alexander (Rice University), T. Barklay (NASA GSFC), T. Bastian (NRAO), S. Boro Saikia (University of Vienna), A. S. Brun (Département d'Astrophysique-AIM, CEA), O. Cohen (University of Massachusetts Lowell), M. Cuntz (UTA), W. Danchi (GSFC/SEEC), J. Davenport (UWA), J. DeNolfo (NASA GSFC), R. DeVore (NASA GSFC), C. F. Dong (Princeton University), J. J. Drake (Harvard-CfA), K. France (CU Boulder), F. Fraschetti (CfA-Harvard/University of Arizona), K. Herbst (University of Kiel), K. Garcia-Sage (GSFC/SEEC/CUA), M. Gillon (University of Liège), A. Glocer (GSFC/SEEC), J. L. Grenfell (German Aerospace Center), G. Gronoff (NASA LaRC/SSAI), N. Gopalswamy (NASA/GSFC), M. Guedel (University of Vienna), H. Hartnett (Arizona State University), H. Harutyunyan (BAO), N. R. Hinkel (Arizona State University), A. G. Jensen (UNK), M. Jin (LMSAL), C. Johnstone (University of Vienna), S. Kahler (AFRL/RVBS), P. Kalas (UC Berkeley), S. R. Kane (UC Riverside), C. Kay (NASA GSFC)/SEEC), I.N. Kitiashvili (BAERI/NASA ARC), O. Kochukhov (Uppsala University), D. Kondrashov (UCLA), J. Lazio (NASA/JPL), J. Leake (GSFC), G. Li (University of Alabama), J. Linsky (CU Boulder)T. Lueftinger (University of Vienna), B. Lynch (UC Berkeley), W. Lyra (Max Planck Institute for Astronomy), A. M. Mandell (GSFC/SEEC), K. E. Mandt ( Johns Hopkins University APL), H. Maehara (NAOJ), M. S. Miesch (JCSDA), A. M. Mickaelian (BAO), S. Mouchou (CfA), Y. Notsu (Kyoto University), L. Ofman (NASA GSFC/CUA), L. D. Oman (GSFC/SEEC), R. A. Osten (STScI, JHU), R. Oran (MIT), R. Petre (NASA GSFC), R. M. Ramirez (ELSI), G. Rau (NASA GSFC), S. Redfield (Wesleyan University), V. Réville (UC Los Angeles), S. Rugheimer (University of Oxford), M. Scheucher (TU Berlin), J. E. Schlieder (NASA GSFC/SEEC), K. Shibata (Kyoto University), J. D. Schnittman (NASA GSFC), David Soderblom (STScI), A. Strugarek (Département d'Astrophysique-AIM, CEA), J. D. Turner (Cornell University), A. Usmanov (University of Delaware), Van Der Holst (University of Michigan), A. Vidotto (University of Dublin), A. Vourlidas (JHU), M.J. Way (NASA/GISS/SEEC), Wolk, G. P. Zank (University of Alabama in Huntsville), P. Zarka (Obs. Paris/CNRS/PSL), R. Kopparapu (NASA GSFC/SEEC), S. Babakhanova (NASA GSFC/SEEC/MIT), A. A. Pevtsov (National Solar Observatory), Y. Lee (NASA GSFC/SEEC/UMBC), W. Henning (Univ. Maryland/NASA GSFC), K. D. Colón (NASA GSFC/SEEC), E. T. Wolf (CU Boulder)

## 1. Introduction.

With over 4000 exoplanets detected by Kepler and other ground-based and space missions, the field of exoplanetary science is making rapid progress both in statistical studies of exoplanet properties as well as in individual characterization. As these missions provide an emerging picture of formation and evolution of exoplanetary systems, the search for habitable worlds becomes one of the fundamental issues to address. To tackle such a complex challenge, we need to specify the conditions favorable for the origin, development and sustainment of life as we know it. This requires the understanding of global (astrospheric) and local (atmospheric, surface and internal) environments of exoplanets in the framework of the physical processes of the interaction between evolving planet-hosting stars along with exoplanetary evolution over geological timescales, and the resulting impact on climate and habitability of exoplanets. Feedbacks between astrophysical, physico-chemical atmospheric and geological processes can only be understood through interdisciplinary studies with the incorporation of progress in heliophysics, astrophysics, planetary, Earth sciences, astrobiology, and the origin of life communities. The assessment of the impacts of host stars on the climate and habitability of terrestrial (exo)planets and potential exomoons around them may significantly modify the extent and the location of the habitable zone (HZ) and provide new directions for searching for signatures of life. Thus, characterization of stellar ionizing outputs in the portion of space between host stars and their exoplanets becomes an important task for further understanding the extent of habitability in the universe.

The goal of this white paper is to identify and describe promising key research goals to aid the theoretical characterization and observational detection of ionizing radiation from quiescent and flaring upper atmospheres of planet hosts as well as properties of stellar coronal mass ejections (CMEs) and stellar energetic particle (SEP) events. This white paper is based on the roadmap paper and a review on exoplanetary space weather (Airapetian + 2019a,b) reflecting recommendations of the NASA NexSS sponsored Exoplanetary Space Weather Workshop and the 1$^{st}$ Sellers Exoplanetary Environment Collaboration (SEEC) Symposium (Airapetian 2018). This paper complements the Astro2020 Decadal Survey on Astronomy & Astrophysics white paper "Magnetic Fields of Extrasolar Planets: Planetary Interiors and Habitability" (Lazio + 2019) and "X-ray Studies of Exoplanets" (Wolk + 2019).

## 2. Key questions related to drivers and signatures of exoplanetary space weather

Solar flares, quiescent winds, CMEs and associated SEP events produce disturbances in interplanetary space - collectively referred to as space weather (SW), which interact with the Earth's upper atmosphere and can cause dramatic impacts on space- and ground-based technological systems (Airapetian + 2019b). In this paper we present the roadmap for the study of various aspects of star-planet interactions in a global exoplanetary system environment with a systematic, integrated approach using theoretical modeling, observational and laboratory methods combining tools and methodologies of four science disciplines: astrophysics, heliophysics, planetary and Earth science as shown in Fig. 1. With the *Kepler* Space Telescope discoveries and characterization of energetic flares (or superflares) on active F-, G-, K-, and M- stars, exoplanetary space weather has emerged as an important subfield of exoplanetary science. This has occurred as a result of research demonstrating the impact of the stellar ionizing radiation fluxes, winds and CMEs on exoplanets within close-in habitable zones (HZs) around active stars. Exoplanetary space weather could therefore have a crucial influence upon habitability.

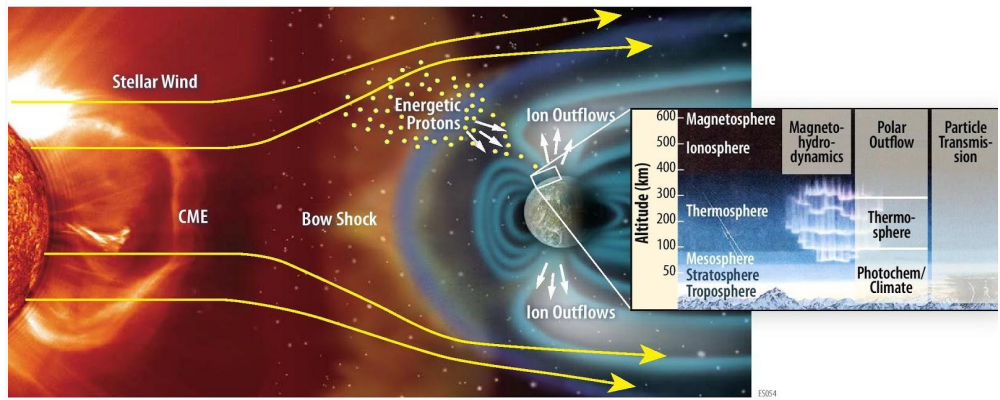

*Figure 1. Schematic view of the complex exoplanetary SW system that incorporates the physical processes driving stellar activity and associated SW including stellar flares, CMEs and their interactions with an exoplanetary atmosphere driven by its internal dynamics. While stellar winds and CMEs affect the shape of an exoplanetary magnetosphere, X-ray and Extreme Ultraviolet Emission (XUV) and energetic particles accelerated by coronal flares or CME-driven shocks enter the atmosphere. Their combined effects drive outflows from the exoplanetary atmosphere. These processes are controlling factors of exoplanetary climate and habitability.*

Stellar activity provides a mechanism by which host stars may have profound effects on the physical and chemical evolution of exoplanetary atmospheres. As stars age, they lose angular momentum via winds (Weber & Davis 1967) and coronal mass ejections (Aarnio+2011), and thus their rotation rates (and Rossby number, the ratio of the rotation rate to the local convective turnover time) change over time. Lower spin rates (higher Rossby numbers) produce weaker magnetic fields (Skumanich 1972; Vidotto+2014) and generate less frequent and lower energy transient magnetic activity (flares, CMEs etc). Thus, the characterization of stellar ages through its proxies becomes one of the central issues in exoplanetary science. The first proxy, the stellar rotation rate, however, suffers from uncertainties associated with young (0.1-0.7 Gyr) stellar clusters implying that the X-ray evolutionary tracks first diverge (i.e., $L_X$ of a slow rotator rapidly decays with time while that of a fast rotator does not) and then converge again after several hundred Myr (Tu+2015). This raises the question as to whether our Sun was a slow, intermediate, or fast rotator during its youth. Thus, realistic theoretical and semi-empirical models of magnetic braking are required to characterize the evolution of stellar angular momentum. The second proxy, photospheric lithium abundance, also varies by a factor of few in active young stars introducing another uncertainty factor in determination of stellar ages (Bouvier+2014). This is especially important given that the duration of high magnetically driven stellar activity, varies with age for F-, G-, K-, and M- dwarfs. For example, if solar (G-type) stars retain their activity for a half a billion years, K dwarfs retain for over 1-2 billion years, while late M dwarfs show a prolonged phase of activity up to 5-8 Gyr (West+2008; France + 2016; Doyle + 2018). Other important proxies of stellar age include metallicity, the level of chromospheric activity measured by Ca II lines (for an example in evolved stars see e.g. Rau et al. 2018) and the level of stellar activity in terms of frequency, magnitude and energies of stellar flares. The Kepler survey suggests a clear correlation between maximum flare energy and stellar activity represented by starspot area. Can this correlation be extended to predict the maximum possible impacts of CMEs and SEPs for individual exoplanets from their host stars (Maehara + 2017; Namekata + 2019)? The associated SW effects represent a vastly underestimated factor regarding the physical impact on atmospheric erosion from CMEs and chemical impacts of energetic particles including SEPs and Galactic Cosmic Rays. Numerous studies of the photochemical response of Earth's middle atmosphere to large SEPs is characterized by strong production of $NO_x$, $HO_x$ constituents and destruction of ozone (see e.g. Grenfell + 2012; Airapetian + 2017a). These studies aim to address to the following four crucial questions: *(1) What types of planet hosting stars can support relatively thick (> 1 mbar) atmospheres of exoplanets in their HZs? (2) What are the strongest biosignatures of an Earth-like biosphere which experiences XUV/SEP fluxes? (3) How significant is ozone destruction in oxygenated exoplanetary atmospheres of Earth-like planets around F-M dwarfs? (4) What are the effects of superflare-induced SEPs on the prebiotic chemistry of rocky exoplanets around G-M dwarfs and at what level can the UV and particle radiation be lethal for surface life forms?*

The evolving magnetic environments of planet-hosting stars determine conditions for exoplanetary systems that are critically important for either promoting or destroying ecological environments of life. Therefore, the characterization of the signatures of star-planet interaction is a high priority topic in upcoming research. This includes their modeling in the context of a global planetary system environment with a systematic, integrated approach using theoretical and observational methods combining tools and methodologies of the four NASA science disciplines: astrophysics, heliophysics, planetary and Earth science. ***Such an approach is timely as high-quality coordinated radio, IR, optical, spectropolarimetric (ZDI maps), FUV/NUV and X-ray observations of planet hosts are required as input and output for a number of sophisticated multi-dimensional physics-based magnetohydrodynamic (MHD), hybrid, kinetic and test-particle models.***

The coordinated multi-wavelength multi-observatory observational effort based on HST, XMM-Newton, Chandra, NICER, Swift, ground-based photometric, spectropolarimetric and spectroscopic observations in the near UV-optical and radio bands is a crucial program for the characterization of energy fluxes from F-, K-, G- and M- dwarfs, because stellar surface magnetic field may vary on time scales of a few days to years (Boro-Saikia +2018; Airapetian+2019b). Such observations for the Sun provide a firm framework for the predictive modeling of the major properties of the solar corona and its expansion into the solar wind, the initiation of solar flares and CME and SEP events and the propagation of SEP through the heliosphere (Jin+2013; Oran+2017; MacNeice+2018; Hu+2018; Fraschetti+19). Recent extensions of data driven heliospheric global corona and wind MHD models to stellar coronal and astrospheric environments include ARMS3D (Lynch + 2019), ALF3D (Airapetian & Usmanov 2016), BATSRUS/AWSoM, PLUTO, EUHFORIA (Réville & Brun 2017; Pomoell & Poedts 2018; Vidotto+2018; Airapetian +2019b) and iPATH (Fu+2019). They were recently applied to the extreme environments of young magnetically active stars to develop the first comprehensive models of global corona and wind systems (Airapetian & Usmanov 2016; Réville+2016; Dong +2017; Alvarado-Gomez+2018). Their outputs include the following:

(1) Stellar XUV flux cannot be reliably determined as it is heavily absorbed by the ISM even for closest stars. However, XUV fluxes are required input parameters for the characterization of photoionization and photodissociation which are crucial for modeling exoplanetary atmospheric dynamics and escape (Ribas+2005; Airapetian+2017a; Garcia-Sage+2017; Johnstone+2018;

Lammer 2018; Glocer+2018). This effort should be combined with empirical and semi-empirical methods which reconstruct XUV fluxes using FUV/NUV and radiative transfer modeling efforts (Linsky+2013; France +2016; Peacock+2018; Richey-Yowell+2019).

(2) The electromechanical flux from stellar winds can only be empirically constrained using magnetohydrodynamic models (Wood+2005). Stellar winds interact with exoplanetary magnetospheres and can perturb them, introducing Joule heating into the upper atmospheres of exoplanets; this in turn promotes atmospheric escape, which can be a negative factor of habitability for close-in planets (Cohen+2014; Airapetian+2017a; Garcia-Sage+2017; Dong+2018; Strugarek 2016; Vidotto +2018).

(3) Multi-dimensional MHD models of stellar superflares and super CMEs from active F-M dwarfs are needed. Recent models of CMEs from the young Sun's proxies suggest that the models can capture the important physics of these eruptive processes on the Sun and other stars (Alvarado-Gomez +2018; Lynch+2019). These models will greatly benefit from statistical studies of starspots and flare occurrence on F-M dwarfs using Kepler, TESS and future PLATO and JWST missions. Studies of the solar activity using realistic simulations have been demonstrated a great potential to understand physical processes behind observed phenomena. Development of 3D radiative MHD modeling of the stellar atmospheric plasma dynamics (Trampedach+2014; Kitiashvili+2016) will help to understand origin and properties of the observed activity phenomena such as starspots and flares, and allow develop new approaches to characterize manifestations of stellar activity.

(4) Reconstruction of fluence and energy distribution of stellar energetic particles accelerated in wind and CME-driven shocks are needed from F-M dwarfs. The precipitation of high energy protons into the lower parts of exoplanetary atmospheres causes enhanced ionization that promotes chemical changes of atmospheric species critical for exoplanetary climate (Tabataba-Vakili + 2016; Scheucher + 2018; Airapetian + 2017b). This process can initiate the rise of prebiotic molecules including HCN on exoplanets (Airapetian + 2016; Ranjan et al. 2018; Rimmer + 2018; Lingam + 2018). Theoretical models should use the empirical correlations between solar SEP and flare events (Kahler 2013; Kahler & Vourlidas 2014; Youngblood+17), extrapolated to young active stars (Fraschetti+19; Fu+19), to provide spectral signatures of biogenic conditions on exoplanets. These outputs (1)-(4) should be modelled for the large array of stellar targets to understand the ranges of physical fluxes on exoplanets at various phases of stellar evolution. This is especially important given that the duration of high magnetically driven stellar activity, the driver of SW, varies with age for F-M dwarfs.

**3. Recommendation**

Understanding the drivers and fluxes of ionizing radiation from F-M type stars hosting exoplanets is critical for characterization of their impact on exoplanetary atmospheres, will be an organic component of exoplanetary science over the next 10-20 years. Given the current progress in developing theoretical modeling and observational tools required for such characterization, we urge the committee to consider the following recommendations:

1. Recognize that modeling of extended astrospheres of stars hosting planets is one of the major components for characterizing habitable exoplanets, especially impacts on atmospheric erosion, chemistry and surface radiation dosages.

2. Promote the development of coordinated multi-wavelength multi-observatory programs to derive critical inputs for theoretical and empirical tools using X-ray, FUV, NUV, radio observations and surface magnetic field maps (magnetograms). The community should strongly encourage the development of new X-ray space telescope (see Wolk et al. 2019), FUV/NUV space telescope (The Star-Planet Activity Research CubeSat (SPARCS), see Scowen et al. 2018; UV space probe mission, CETUS, Danchi et al. 2018, Colorado Ultraviolet Transit Experiment (CUTE), Fleming + 2018) missions, and radio telescope facilities (LOFAR and Very large Array (ngVLA) (Osten et al. 2018) to open new windows into the nature of nearby planet hosting stars.

3. Perform direct, detailed characterization of stellar magnetic fields, in particular those of key exoplanet host stars, through high-resolution spectroscopy, spectropolarimetry and interferometric polarimetry using data obtained with ground-based optical and infrared facilities including the upcoming instrument SPIRou for infrared spectropolarimetry at the CFHT. Characterize the evolution of stellar magnetic structures in large samples of stars using indirect proxies, including spots and their association with flares using data from Kepler, TESS, and upcoming missions including CHEOPS, JWST, PLATO 2.0 and ARIEL. Develop dynamo models of F-M stars. Also, we encourage the extension of heliospheric models to astrospheric models to study initiation and development of stellar flares & CMEs. These models can also be used to characterize the magnetic connectivity between a star and a planet. This is important for SW impact on habitability as it will determine (for instance) the trajectories of stellar energetic particles. These models will should also characterize the impact of activity transmission spectra of exoplanets, which is an unexplored area.

4. Refine the characterization of the ages of planet hosting stars using Li, rotation rates, CaII H&K, and patterns of magnetic activity in the form of frequency distribution of stellar flares, their duration and maximum energy and maximal sizes of starspots. Thus, dedicated observations and comprehensive characterization of flares at different phases of evolution of F-M stars are required along with flare frequency.

5. Understand the SW impact on exoplanetary atmospheres of diverse chemistries and atmospheric escape. This includes development of multi-fluid hydrodynamic, MHD, hybrid and particle-in-cell models of atmospheres of terrestrial and gas giant type exoplanets that incorporate the range from magnetosphere-ionosphere to lower atmospheric layers. Such comprehensive models should characterize the conditions at which various mechanisms of atmospheric erosion including hydrodynamic escape, ion and neutral escape become important factors contributing to atmospheric pressure, and thus, would be crucial for habitability.

6. Understand SW impact on the atmosphere is vital to also understanding the surface dosage of ionizing radiation as well as the planet's climate, habitability, and the detection and interpretation of novel biosignatures.

7. Characterize the impact of SE from the young Sun on atmospheric chemistry and erosion of early Venus, Earth and Mars as "exoplanets" with constraints available from isotopic studies. Specifically, combined theoretical and laboratory studies of prebiotic chemistry impacted by the intense solar activity of the young Sun and pathways to formation of complex biology of life should be a high priority for the next 10 years.

8. Characterize exoplanetary prebiotic chemical pathways to life induced by the SW factors from F-M dwarfs in laboratory experiments using stellar outputs in the form of UV and energetic particles. These experiments will constrain calibrated chemical species serving as intermediary molecules and ions on the pathway to the initiation and development of primitive life.